\documentstyle[12pt]{article}
\begin{document}
\title{Two/Three-Flavor Oscillation and \\ MSW/Vacuum Oscillation Solution of Neutrinos \\ 
in the SO(3) Gauge Model}
\author{Yue-Liang  Wu\footnote{ylwu@ITP.AC.CN}  \\  
\\
Institute of Theoretical Physics, Chinese Academy of Sciences, \\
 P.O. Box 2735, Beijing 100080, China } 
\date{AS-ITP-99-04}
\maketitle

\begin{abstract} 
  Three interesting scenarios for neutrino mixing, i.e., (i) small-large mixing scenario, 
(ii) nearly bi-maximal mixing scenario and (iii) three-flavor oscillation scenario, are analyzed 
in connection with three possible assignments of the maximal CP-violating phase after 
spontaneous symmetry breaking of SO(3) in the model with gauged SO(3) lepton flavor symmetry.  
As a consequence, it is found that the scenario (ii) is more 
reliable to be constructed to reconcile both solar and atmospheric neutrino data. 
Though three Majorana neutrino masses in all scenarios 
can be nearly degenerate, in the scenarios (ii) and (iii) masses of the neutrinos are 
allowed to be large enough to play a significant cosmological role and in the scenario (i) 
the fraction $\Omega_{\nu}/\Omega_{m}$ is bounded to be $\Omega_{\nu}/\Omega_{m} < (4-10)\%$ 
for $\Omega_{m} = (1.0-0.4)$ and $h=0.6$.   
\end{abstract}
\newpage

\section{Introduction}

It has been shown that the current neutrino data on both atmospheric\cite{SUPERK1} and 
solar\cite{SUPERK2, SOLAR}  neutrino flux anomalies (the solar neutrino flux measured recently 
by the Super-Kamiokande collaboration is only about half of that expected from the `BP' standard
solar model (`BP' SSM)\cite{BP}) can be explained by a nearly `bi-maximal' mixing pattern 
(that include the bi-maximal mixing pattern\cite{BMAX} and 
democratic mixing pattern\cite{DEM}). In the recent papers\cite{YLWU}, we have investigated
gauged SO(3) symmetry for three lepton families and derived a realistic scenario for such patterns. 
In this paper, we consider three possible 
scenarios for neutrino mixing: (i) small-large mixing scenario and 
(ii)  nearly bi-maximal mixing scenario as well as 
(iii) three-flavor oscillation scenario due to
three possible assignments of the maximal CP-violating phase after spontaneous symmetry breaking 
of SO(3). The SO(3) symmetry can naturally lead three Majorana neutrino masses in all scenarios 
to be nearly degenerate\cite{CM,MA,CW,BHKR}. 
In the scenarios (ii) and (iii), masses of the neutrinos are allowed to be large enough to play a 
significant cosmological role and in the scenario (i) they 
are constrained by the neutrinoless double beta decay. In particular, it is seen that the scenario (ii) 
is more reliable to be constructed to reconcile both solar and atmospheric neutrino data. 
Our paper is organized as follows: in the section 2, 
we provide a description of the model based on the SO(3) gauge symmetry of lepton flavor. The small-large 
mixing scenario is discussed in the section 3. In the section 4, the nearly bi-maximal mixing 
scenario is analysed. The three-flavor oscillation scenario is discussed in the section 5.  
Conclusions and remarks are presented in the last section. 

\section{ The Model} 

  We begin with the following SO(3)$_{F}\times$SU(2)$_{L}\times$U(1)$_{Y}$ 
invariant effective lagrangian for leptons 
\begin{eqnarray}
{\cal L} & = &  \frac{1}{2}g'_{3}A_{\mu}^{k}
\left( \bar{L}_{i}\gamma^{\mu} (t^{k})_{ij}L_{j} 
+ \bar{e}_{R i} \gamma^{\mu}(t^{k})_{ij}e_{R j} \right) 
+ D_{\mu}\varphi^{\ast}D^{\mu}\varphi + D_{\mu}\varphi^{'\ast}D^{\mu}\varphi' \nonumber \\
& + & \left(C_{1}\frac{\varphi_{i}\varphi_{j}}{M_{1}M_{2}} + 
C'_{1}\frac{\varphi'_{i}\varphi'_{j}}{M'_{1}M'_{2}}\frac{\chi}{M} + C''_{1} 
\frac{\chi'}{M'}\delta_{ij} \right)\bar{L}_{i} \phi_{1}e_{R\ j} + h.c.
 \\
& + & \left(C_{0}\delta_{ij} + C'_{0}\frac{\varphi_{i}
\varphi_{j}^{\ast}}{M_{2}^{2}} + C''_{0}\frac{\varphi'_{i}
\varphi_{j}^{'\ast}}{M_{2}^{'2}} \right)\frac{1}{M_{N}} 
\bar{L}_{i} \phi_{2}\phi_{2}^{T}L_{j}^{c} + h.c. + {\cal L}_{SM} \nonumber 
\end{eqnarray}
which is assumed to be resulted from integrating out heavy particles. 
Where ${\cal L}_{SM} $ denotes the lagrangian of the standard 
model. $\bar{L}_{i}(x) = (\bar{\nu}_{i}, \bar{e}_{i})_{L}$ 
(i=1,2,3) are the SU(2)$_{L}$ doublet leptons. $e_{R\ i}$ (i=1,2,3) are 
the three right-handed charged leptons. $\phi_{1}(x)$ and $\phi_{2}(x)$ 
are two Higgs doublets. $\varphi^{T}=(\varphi_{1}(x), \varphi_{2}(x), 
\varphi_{3}(x))$ and $\varphi'^{T} = (\varphi'_{1}(x), \varphi'_{2}(x), 
\varphi'_{3}(x))$ are two complex SO(3) triplet scalars. $\chi(x)$ and 
$\chi'(x)$ are two singlet scalars. $M_{1}$, $M_{2}$, $M$, 
$M'_{1}$, $M'_{2}$, $M'$ and $M_{N}$ are possible mass scales concerning 
heavy fermions. $C_{a}$, $C'_{a}$ and 
$C''_{a}$ ($a=0,1$) are six coupling constants. The structure of the above 
effective lagrangian can be obtained by imposing an additional U(1) symmetry, 
which is analogous to the construction of the $C_{0}$ and $C_{1}$ terms 
discussed in detail in ref.\cite{YLWU}. After the symmetry 
SO(3)$_{F}\times$SU(2)$_{L}\times$U(1)$_{Y}$ is broken down to the U(1)$_{em}$ 
symmetry, we obtain mass matrices of the neutrinos and charged leptons as follows
\begin{eqnarray}
(M_{e})_{ij} & = & m_{1} \frac{\hat{\sigma}_{i}\hat{\sigma}_{j}}{\sigma^{2}} 
+ m'_{1} \frac{\hat{\sigma}'_{i}\hat{\sigma}'_{j}}{\sigma^{'2}} + m''_{1} 
\delta_{ij} \nonumber \\
(M_{\nu})_{ij} & = &  m_{0} \delta_{ij} + m'_{0} \frac{\hat{\sigma}_{i}
\hat{\sigma}_{j}^{\ast} + \hat{\sigma}_{j}\hat{\sigma}_{i}^{\ast} }{2\sigma^{2}} 
+ m''_{0} \frac{\hat{\sigma}'_{i}\hat{\sigma}_{j}^{'\ast} + 
\hat{\sigma}'_{j}\hat{\sigma}_{i}^{'\ast}}{2\sigma^{'2}}  
\end{eqnarray}
where the mass matrices $M_{e}$ and $M_{\nu}$ are defined in the basis
 ${\cal L}_{M} = \bar{e}_{L}M_{e}e_{R} + 
\bar{\nu}_{L}M_{\nu}\nu^{c}_{L} + h.c. $. The constants $\hat{\sigma}_{i} = 
<\varphi_{i}(x)>$ and $\hat{\sigma}'_{i} = <\varphi'_{i}(x)>$ 
represent the vacuum expectation values of the two triplet 
scalars $\varphi(x)$ and $\varphi'(x)$. The six mass parameters are defined as:
$m_{0} = C_{0}v_{2}^{2}/M_{N}$, $m'_{0} = C'_{0}(\sigma^{2}/M_{2}^{2})(v_{2}^{2}/M_{N})$, 
$m''_{0} = C''_{0}(\sigma^{'2}/M_{2}^{'2})(v_{2}^{2}/M_{N})$,
$m_{1} = C_{1}v_{1}\sigma^{2}/M_{1}M_{2}$, $m'_{1} = C'_{1}
(\xi/M)(v_{1}\sigma^{2}/M'_{1}M'_{2})$ and $m''_{1} = C''_{1}v_{1}\xi'/M$. Here 
$\sigma = \sqrt{|\hat{\sigma}_{1}|^{2} + |\hat{\sigma}_{2}|^{2} + 
|\hat{\sigma}_{3}|^{2}}$ and $\sigma' = \sqrt{|\hat{\sigma}'_{1}|^{2} + 
|\hat{\sigma}'_{2}|^{2} + |\hat{\sigma}'_{3}|^{2}}$. 
$\xi =<\chi(x)>$ and $\xi'=<\chi'(x)>$ denote the vacuum expectation values of 
the two singlet scalars. 

  Utilizing the gauge symmetry property, it is convenient to reexpress 
the complex triplet scalar fields $\varphi_{i}(x)$ and $\varphi'_{i}(x)$ 
in terms of the SO(3) rotational fields $O(x)= e^{i\eta_{i}(x)t^{i}},\  
O'(x)= e^{i\eta'_{i}(x)t^{i}} \in $ SO(3). In general, there exist three different vacuum 
structures in connection with three possible assignments of the imaginary amplitude field
(for completeness and comparison, we will also include the case discussed in ref. \cite{YLWU})
\begin{eqnarray} 
& A: & \left( \begin{array}{c}
  \varphi_{1}(x) \\
  \varphi_{2}(x)   \\
  \varphi_{3}(x)   \\
\end{array} \right) = \frac{e^{i\eta_{i}(x)t^{i}}}{\sqrt{2}}
\left( \begin{array}{c}
  \rho_{1}(x) \\
  \rho_{2}(x)   \\
  i\rho_{3}(x)   \\
\end{array} \right); \qquad \left( \begin{array}{c}
  \varphi'_{1}(x) \\
  \varphi'_{2}(x)   \\
  \varphi'_{3}(x)   \\
\end{array} \right) = \frac{e^{i\eta'_{i}(x)t^{i}}}{\sqrt{2}}
\left( \begin{array}{c}
  \rho'_{1}(x) \\
  \rho'_{2}(x)   \\
  i\rho'_{3}(x)   \\
\end{array} \right)   \nonumber \\
& B: & \left( \begin{array}{c}
  \varphi_{1}(x) \\
  \varphi_{2}(x)   \\
  \varphi_{3}(x)   \\
\end{array} \right) = \frac{e^{i\eta_{i}(x)t^{i}}}{\sqrt{2}}
\left( \begin{array}{c}
  \rho_{1}(x) \\
  i\rho_{2}(x)   \\
  \rho_{3}(x)   \\
\end{array} \right); \qquad \left( \begin{array}{c}
  \varphi'_{1}(x) \\
  \varphi'_{2}(x)   \\
  \varphi'_{3}(x)   \\
\end{array} \right) = \frac{e^{i\eta'_{i}(x)t^{i}}}{\sqrt{2}}
\left( \begin{array}{c}
  \rho'_{1}(x) \\
  i\rho'_{2}(x)   \\
  \rho'_{3}(x)   \\
\end{array} \right)  \\
& C: & \left( \begin{array}{c}
  \varphi_{1}(x) \\
  \varphi_{2}(x)   \\
  \varphi_{3}(x)   \\
\end{array} \right) = \frac{e^{i\eta_{i}(x)t^{i}}}{\sqrt{2}}
\left( \begin{array}{c}
  i\rho_{1}(x) \\
  \rho_{2}(x)   \\
  \rho_{3}(x)   \\
\end{array} \right); \qquad \left( \begin{array}{c}
  \varphi'_{1}(x) \\
  \varphi'_{2}(x)   \\
  \varphi'_{3}(x)   \\
\end{array} \right) = \frac{e^{i\eta'_{i}(x)t^{i}}}{\sqrt{2}}
\left( \begin{array}{c}
  i\rho'_{1}(x) \\
  \rho'_{2}(x)   \\
  \rho'_{3}(x)   \\
\end{array} \right)  \nonumber 
\end{eqnarray}
Here the three rotational fields $\eta_{i}(x)$ ($\eta'_{i}(x)$) and 
the three amplitude fields $\rho_{i}(x)$ ($\rho'_{i}(x)$) reparameterize 
the six real fields of the complex triplet scalar field $\varphi(x)$ 
($\varphi(x)$). SO(3) gauge symmetry allows one to remove three degrees 
of freedom from the six rotational fields. Thus the vacuum structure of the 
SO(3) symmetry is expected to be determined only by nine degrees of freedom.
These nine degrees of freedom can be taken as $\rho_{i}(x)$, $\rho'_{i}(x)$ and 
$(\eta_{i}(x) - \eta'_{i}(x))$ without lossing generality. Here we will consider 
the following vacuum structure for the SO(3) symmetry breaking
\begin{equation}
<\rho_{i}(x)> = \sigma_{i}, \qquad <\rho'_{i}(x)> = \sigma'_{i}, \qquad 
<(\eta_{i}(x) - \eta'_{i}(x))> = 0
\end{equation}
With this vacuum structure, the mass matrices of the neutrinos and charged 
leptons can be reexpressed as 
\begin{eqnarray}
A: \qquad M_{e} & = & m_{1}\left( \begin{array}{ccc}
  s_{1}^{2}s_{2}^{2} & c_{1}s_{1}s_{2}^{2} & is_{1}c_{2}s_{2}  \\
   c_{1}s_{1}s_{2}^{2} & c_{1}^{2}s_{2}^{2} &  ic_{1}c_{2}s_{2} \\
  is_{1}c_{2}s_{2} & ic_{1}c_{2}s_{2}
 & -c_{2}^{2}  \\ 
\end{array} \right) \nonumber \\
& + & m'_{1}\left( \begin{array}{ccc}
  s_{1}^{'2}s_{2}^{'2} & c'_{1}s'_{1}s_{2}^{'2} & is'_{1}c'_{2}s'_{2}  \\
   c'_{1}s'_{1}s_{2}^{'2} & c_{1}^{'2}s_{2}^{'2} &  ic'_{1}c'_{2}s'_{2} \\
  is'_{1}c'_{2}s'_{2} & ic'_{1}c'_{2}s'_{2}
 & -c_{2}^{'2}  \\ 
\end{array} \right) + m''_{1}\left( \begin{array}{ccc}
  1 & 0 & 0  \\
   0 & 1 & 0 \\
 0 & 0 &  1 \\ 
\end{array} \right) \nonumber \\
B: \qquad M_{e} & = & m_{1}\left( \begin{array}{ccc}
  s_{1}^{2}s_{2}^{2} & ic_{1}s_{1}s_{2}^{2} & s_{1}c_{2}s_{2}  \\
   ic_{1}s_{1}s_{2}^{2} & -c_{1}^{2}s_{2}^{2} &  ic_{1}c_{2}s_{2} \\
  s_{1}c_{2}s_{2} & ic_{1}c_{2}s_{2}
 & c_{2}^{2} \\ 
\end{array} \right) \nonumber \\
& + & m'_{1}\left( \begin{array}{ccc}
  s_{1}^{'2}s_{2}^{'2} & ic'_{1}s'_{1}s_{2}^{'2} & s'_{1}c'_{2}s'_{2}  \\
   ic'_{1}s'_{1}s_{2}^{'2} & -c_{1}^{'2}s_{2}^{'2} &  ic'_{1}c'_{2}s'_{2} \\
  s'_{1}c'_{2}s'_{2} & ic'_{1}c'_{2}s'_{2} & c_{2}^{'2}  \\ 
\end{array} \right) + m''_{1}\left( \begin{array}{ccc}
  1 & 0 & 0  \\
   0 & 1 & 0 \\
 0 & 0 &  1 \\ 
\end{array} \right) \\
C: \qquad M_{e} & = & m_{1}\left( \begin{array}{ccc}
  -s_{1}^{2}s_{2}^{2} & ic_{1}s_{1}s_{2}^{2} & is_{1}c_{2}s_{2}  \\
   ic_{1}s_{1}s_{2}^{2} & c_{1}^{2}s_{2}^{2} &  c_{1}c_{2}s_{2} \\
  is_{1}c_{2}s_{2} & c_{1}c_{2}s_{2} & c_{2}^{2}  \\ 
\end{array} \right) \nonumber \\
& + & m'_{1}\left( \begin{array}{ccc}
  -s_{1}^{'2}s_{2}^{'2} & ic'_{1}s'_{1}s_{2}^{'2} & is'_{1}c'_{2}s'_{2}  \\
   ic'_{1}s'_{1}s_{2}^{'2} & c_{1}^{'2}s_{2}^{'2} &  c'_{1}c'_{2}s'_{2} \\
  is'_{1}c'_{2}s'_{2} & c'_{1}c'_{2}s'_{2} & c_{2}^{'2}  \\ 
\end{array} \right) + m''_{1}\left( \begin{array}{ccc}
  1 & 0 & 0  \\
   0 & 1 & 0 \\
 0 & 0 &  1 \\ 
\end{array} \right) \nonumber
\end{eqnarray}
and 
\begin{eqnarray}
A: \qquad M_{\nu} & = & m_{0}\left( \begin{array}{ccc}
  1 & 0 & 0  \\
   0 & 1 & 0 \\
 0 & 0 &  1 \\ 
\end{array} \right) + 
m'_{0}\left( \begin{array}{ccc}
  s_{1}^{2}s_{2}^{2} & s_{1}c_{1}s_{2}^{2}  & 0 \\
  s_{1}c_{1}s_{2}^{2} & c_{1}^{2}s_{2}^{2} &  0 \\
  0 & 0 & c_{2}^{2} \\ 
\end{array} \right) \nonumber \\
& + & m''_{0}\left( \begin{array}{ccc}
  s_{1}^{'2}s_{2}^{'2} & s'_{1}c'_{1}s_{2}^{'2} & 0  \\
  s'_{1}c'_{1}s_{2}^{'2} & c_{1}^{'2}s_{2}^{'2} &  0 \\
  0 & 0 & c_{2}^{'2} \\ 
\end{array} \right) \nonumber \\
B:\qquad  M_{\nu} & = & m_{0}\left( \begin{array}{ccc}
  1 & 0 & 0  \\
   0 & 1 & 0 \\
 0 & 0 &  1 \\ 
\end{array} \right) + 
m'_{0}\left( \begin{array}{ccc}
  s_{1}^{2}s_{2}^{2} & 0 & s_{1}c_{2}s_{2}  \\
  0 & c_{1}^{2}s_{2}^{2} &  0 \\
  s_{1}c_{2}s_{2} & 0
 & c_{2}^{2}  \\ 
\end{array} \right) \nonumber \\
& + & m''_{0}\left( \begin{array}{ccc}
  s_{1}^{'2}s_{2}^{'2} & 0 & s'_{1}c'_{2}s'_{2}  \\
  0 & c_{1}^{'2}s_{2}^{'2} &  0 \\
  s'_{1}c'_{2}s'_{2} & 0
 & c_{2}^{'2} \\ 
\end{array} \right) \\
C:\qquad   M_{\nu} & = & m_{0}\left( \begin{array}{ccc}
  1 & 0 & 0  \\
   0 & 1 & 0 \\
 0 & 0 &  1 \\ 
\end{array} \right) + 
m'_{0}\left( \begin{array}{ccc}
  s_{1}^{2}s_{2}^{2} & 0 & 0 \\
  0 & c_{1}^{2}s_{2}^{2} &  c_{1}c_{2}s_{2}  \\
  0 & c_{1}c_{2}s_{2}  & c_{2}^{2}  \\ 
\end{array} \right) \nonumber \\
& + & m''_{0}\left( \begin{array}{ccc}
  s_{1}^{'2}s_{2}^{'2} & 0 & 0  \\
  0 & c_{1}^{'2}s_{2}^{'2} &  c'_{1}c'_{2}s'_{2} \\
  0 & c'_{1}c'_{2}s'_{2} & c_{2}^{'2} \\ 
\end{array} \right) \nonumber 
\end{eqnarray}
which correspond to the three vacuum structures $A$, $B$ and $C$ in eq.(3).
Where $s_{1} = \sin \theta_{1} = \sigma_{1}/\sigma_{12}$ and 
$s_{2} = \sin \theta_{2} = \sigma_{12}/\sigma$ 
with $\sigma_{12} = \sqrt{\sigma_{1}^{2} + \sigma_{2}^{2}}$ and 
$\sigma =\sqrt{\sigma_{12}^{2} + \sigma_{3}^{2}}$. Similar definitions 
are for $s'_{1}$ and $s'_{2}$. 

 Note that the two non-diagonal matrices in the mass matrix $M_{e}$ are 
rank one matrices. While it is interesting to observe that when 
the four angles $\theta_{1}$, $\theta_{2}$, $\theta'_{1}$ and $\theta'_{2}$
satisfy the following conditions
\begin{equation} 
\frac{s_{1}}{c_{1}} = \frac{s'_{1}}{c'_{1}}, \qquad \frac{c_{2}}{s_{2}} 
= -\frac{s'_{2}}{c'_{2}}
\end{equation}
which is equivalent to $\sigma'_{1}/\sigma'_{2} 
= \sigma_{1}/\sigma_{2}$, $\sigma'_{12}/\sigma'_{3}=-\sigma_{3}/\sigma_{12}$, the 
two non-diagonal matrices in the mass matrix $M_{e}$ can be simultaneously diagonalized 
by a unitary matrix $U_{e}$ via $M'_{e} = U_{e}^{\dagger} M_{e} U_{e}^{\ast}$. Here
\begin{equation} 
M'_{e}  = \left( \begin{array}{ccc}
  0 & 0 & 0  \\
  0 & m'_{1}& 0  \\
  0 & 0 & m_{1}   \\
\end{array} \right) + m''_{1}U_{e}^{\dagger}U_{e}^{\ast}
\end{equation}
and  
\begin{eqnarray}
A: \qquad U_{e}^{\dagger} & = & \left( \begin{array}{ccc}
  c_{1} & -s_{1} & 0  \\
 ic_{2}s_{1} & ic_{1}c_{2} & -s_{2} \\
 s_{1} s_{2} & c_{1}s_{2} & -ic_{2}  \\
\end{array} \right) \nonumber \\
B: \qquad U_{e}^{\dagger} & = & \left( \begin{array}{ccc}
  ic_{1} & -s_{1} & 0  \\
 c_{2}s_{1} & -ic_{1}c_{2} & -s_{2} \\
 s_{1} s_{2} & -ic_{1}s_{2} & c_{2}  \\
\end{array} \right) \\
C: \qquad U_{e}^{\dagger} & = & \left( \begin{array}{ccc}
  c_{1} & -is_{1} & 0  \\
 -ic_{2}s_{1} & c_{1}c_{2} & -s_{2} \\
 -is_{1} s_{2} & c_{1}s_{2} & c_{2}  \\
\end{array} \right) \nonumber 
\end{eqnarray}
where $U_{e}^{\dagger}U_{e}^{\ast}$ has the following explicit form 
\begin{eqnarray}
A:\qquad U_{e}^{\dagger}U_{e}^{\ast} & = & \left( \begin{array}{ccc}
  1 & 0 & 0  \\
  0 & s_{2}^{2}-c_{2}^{2} & 2ic_{2}s_{2}  \\
  0 & 2ic_{2}s_{2}
 & s_{2}^{2}-c_{2}^{2}  \\ 
\end{array} \right) \nonumber \\
B:\qquad U_{e}^{\dagger}U_{e}^{\ast} & = & \left( \begin{array}{ccc}
  s_{1}^{2}-c_{1}^{2} & 2ic_{1}s_{1}c_{2} & 2ic_{1}s_{1}s_{2}  \\
   2ic_{1}s_{1}c_{2} & c_{2}^{2}(s_{1}^{2}-c_{1}^{2})+s_{2}^{2}
 & c_{2}s_{2}(s_{1}^{2}-c_{1}^{2}) - c_{2}s_{2}  \\
  2ic_{1}s_{1}s_{2} & c_{2}s_{2}(s_{1}^{2}-c_{1}^{2}) - c_{2}s_{2}
 & s_{2}^{2}(s_{1}^{2}-c_{1}^{2})+ c_{2}^{2}  \\ 
\end{array} \right) \\
C:\qquad U_{e}^{\dagger}U_{e}^{\ast} & = & \left( \begin{array}{ccc}
  c_{1}^{2}-s_{1}^{2} & -2ic_{1}s_{1}c_{2} & -2ic_{1}s_{1}s_{2}  \\
   -2ic_{1}s_{1}c_{2} & c_{2}^{2}(c_{1}^{2}-s_{1}^{2})+s_{2}^{2}
 & c_{2}s_{2}(s_{1}^{2}-c_{1}^{2}) - c_{2}s_{2}  \\
  -2ic_{1}s_{1}s_{2} & c_{2}s_{2}(c_{1}^{2}-s_{1}^{2}) - c_{2}s_{2}
 & s_{2}^{2}(c_{1}^{2}-s_{1}^{2})+ c_{2}^{2}  \\ 
\end{array} \right) \nonumber 
\end{eqnarray}
Note that the case B is dual to case C via $c_{1} \leftrightarrow - s_{1}$. 
The hierarchical structure of the charged lepton mass implies that 
$m''_{1} << m'_{1} << m_{1}$, it is then not difficult to see that 
the matrix $M'_{e}$ will be further diagonalized by a unitary matrix 
$U'_{e}$ via $D_{e} = U_{e}^{'\dagger} M'_{e} U_{e}^{'\ast} =  
U_{e}^{'\dagger} U_{e}^{\dagger} M_{e} U_{e}^{\ast} U_{e}^{'\ast}$ with 
\begin{equation} 
D_{e}  = \left( \begin{array}{ccc}
  m_{e} & 0 & 0  \\
  0 & m_{\mu}& 0  \\
  0 & 0 & m_{\tau}   \\
\end{array} \right) 
\end{equation}
and 
\begin{eqnarray}
A: \qquad U_{e}^{'\dagger} & \simeq & \left( \begin{array}{ccc}
  1  & 0 & 0 \\
 0 & 1  & iO(m''_{1}/m_{1}) \\
 0 & iO(m''_{1}/m_{1}) & 1   \\
\end{array} \right) \nonumber \\
B: \qquad U_{e}^{'\dagger} & \simeq & \left( \begin{array}{ccc}
  1 + O(m''_{1}/m'_{1}) & iO(m''_{1}/m'_{1}) & iO(m''_{1}/m_{1})  \\
 iO(m''_{1}/m'_{1}) & 1 + O(m''_{1}/m'_{1}) & O(m''_{1}/m_{1}) \\
 iO(m''_{1}/m_{1}) & O(m''_{1}/m_{1}) & 1 + O(m''_{1}/m_{1}) \\
\end{array} \right) \\
C: \qquad U_{e}^{'\dagger} & \simeq & \left( \begin{array}{ccc}
  1 + O(m''_{1}/m'_{1}) & -iO(m''_{1}/m'_{1}) & -iO(m''_{1}/m_{1})  \\
 -iO(m''_{1}/m'_{1}) & 1 + O(m''_{1}/m'_{1}) & O(m''_{1}/m_{1}) \\
 -iO(m''_{1}/m_{1}) & O(m''_{1}/m_{1}) & 1 +O(m''_{1}/m_{1})  \\
\end{array} \right) \nonumber 
\end{eqnarray}
where $m_{e}= O(m''_{1})$, $m_{\mu}= m'_{1} + O(m''_{1}) $ and 
$m_{\tau}= m_{1} + O(m''_{1})$ define the three charged lepton masses. 
This indicates that the unitary matrix $U'_{e}$ does not significantly 
differ from the unit matrix. Applying the same conditions given in eq.(7), 
the three neutrino mass matrices can be rewritten as
\begin{eqnarray}
& & A:\qquad  M_{\nu} = m_{0}\left( \begin{array}{ccc}
  1 + \Delta_{-}s_{1}^{2} & \Delta_{-}s_{1}c_{1}  & 0 \\
  \Delta_{-}s_{1}c_{1}  & 1 + \Delta_{-}c_{1}^{2} &  0 \\
  0 & 0 & 1 + \Delta_{+}  \\ 
\end{array} \right)  \nonumber \\
& & B:\qquad M_{\nu} = m_{0}\left( \begin{array}{ccc}
  1 + \Delta_{-}s_{1}^{2} & 0 & 2\delta_{-}s_{2}c_{2}s_{1}  \\
  0 & 1 + \Delta_{-}c_{1}^{2} &  0 \\
  2\delta_{-}s_{2}c_{2}s_{1} & 0
 & 1 + \Delta_{+}  \\ 
\end{array} \right)  \\
& & C:\qquad M_{\nu} = m_{0}\left( \begin{array}{ccc}
  1 + \Delta_{-}s_{1}^{2} & 0 & 0 \\
  0 & 1 + \Delta_{-}c_{1}^{2} &  2\delta_{-}s_{2}c_{2}c_{1}  \\
  0 & 2\delta_{-}s_{2}c_{2}c_{1}  & 1 + \Delta_{+}  \\ 
\end{array} \right)  \nonumber
\end{eqnarray}
with 
\begin{equation}
\Delta_{\pm} = \delta_{+} \pm \delta_{-}\cos 2\theta_{2}, 
\qquad \delta_{\pm} = (m'_{0}\pm m''_{0})/2m_{0}
\end{equation}
The three type neutrino mass matrices can be easily diagonalized by the orthogonal matrix
$O_{\nu}$ via $O_{\nu}^{T}M_{\nu}O_{\nu}$. Explicitly, the matrix $O_{\nu}$ has the following forms 
for the cases $A$, $B$ and $C$ 
\begin{eqnarray}
A:\qquad O_{\nu} & = & \left( \begin{array}{ccc}
  c_{\nu} & s_{\nu}  & 0 \\
 - s_{\nu} & c_{\nu} & 0 \\
 0 & 0 & 1  \\
\end{array} \right) \nonumber \\
B:\qquad O_{\nu} & = & \left( \begin{array}{ccc}
  c_{\nu} & 0 & s_{\nu}  \\
 0 & 1 & 0 \\
 - s_{\nu} & 0 & c_{\nu}  \\
\end{array} \right) \\
C:\qquad O_{\nu} & = & \left( \begin{array}{ccc}
  1 & 0 & 0  \\
 0 & c_{\nu} & s_{\nu} \\
 0 & - s_{\nu} & c_{\nu}  \\
\end{array} \right) \nonumber 
\end{eqnarray}
with 
\begin{eqnarray}
A: \qquad \tan2\theta_{\nu} & = & \tan2\theta_{1} \nonumber \\
B: \qquad \tan2\theta_{\nu} & = & 2\delta_{-}s_{1}\sin2\theta_{2}/(\Delta_{+} -\Delta_{-}s_{1}^{2})  \\
C: \qquad \tan2\theta_{\nu} & = & 2\delta_{-}c_{1}\sin2\theta_{2}/(\Delta_{+} -\Delta_{-}c_{1}^{2}) \nonumber 
\end{eqnarray}
When going to the physical mass basis of the neutrinos and charged leptons, 
we then obtain the CKM-type lepton mixing matrix $U_{LEP}$  
appearing in the interactions of the charged weak gauge bosons and leptons, i.e.,
${\cal L}_{W} = \bar{e}_{L}\gamma^{\mu}U_{LEP} \nu_{L} W_{\mu}^{-} + h.c. $. 
Expilcitly, we have for the case A
\begin{eqnarray}
A:\qquad U_{LEP} & = & U_{e}^{'\dagger}U_{e}^{\dagger}O_{\nu} = U_{e}^{'\dagger}\left( \begin{array}{ccc}
  1 & 0 & 0  \\
 0 & ic_{2} & -s_{2} \\
 0 & s_{2} & -ic_{2}  \\
\end{array} \right) \nonumber \\
 & \simeq & \left( \begin{array}{ccc}
  1  & 0 & 0 \\
 0 & ic_{2} & -s_{2} \\
 0 & s_{2} & -ic_{2}   \\
\end{array} \right)
\end{eqnarray}
 Where we have neglected the relative small terms of order $O(m''_{1}/m_{1})$.
The three neutrino masses are found to be
\begin{eqnarray}
m_{\nu_{e}} & = &  m_{0}
\nonumber \\
m_{\nu_{\mu}} & = &  m_{0}[ 1 + \Delta_{-} ]   \\
m_{\nu_{\tau}} & = &  m_{0}[ 1 + \Delta_{+} ] \nonumber 
\end{eqnarray}
For the case B, the lepton mixing matrix and three neutrino masses are given by 
\begin{equation}
B:\qquad U_{LEP} = U_{e}^{'\dagger}U_{e}^{\dagger}O_{\nu} = U_{e}^{'\dagger}\left( \begin{array}{ccc}
  ic_{1}c_{\nu} & -s_{1} & ic_{1}s_{\nu}  \\
 c_{2}s_{1}c_{\nu}+ s_{2}s_{\nu} & -ic_{1}c_{2} & c_{2}s_{1}s_{\nu}-s_{2}c_{\nu} \\
 s_{1} s_{2}c_{\nu}-c_{2}s_{\nu} & -ic_{1}s_{2} & s_{1}s_{2}s_{\nu}+c_{2}c_{\nu}  \\
\end{array} \right)
\end{equation}
and
\begin{eqnarray}
m_{\nu_{e}} & = &  m_{0}[ 1 + \frac{1}{2}(\Delta_{+} + \Delta_{-}s_{1}^{2} ) - 
\frac{1}{2} (\Delta_{+} - \Delta_{-}s_{1}^{2})\sqrt{1 + \tan^{2}2\theta_{\nu} } \  ] 
\nonumber \\
m_{\nu_{\mu}} & = &  m_{0}[ 1 + \Delta_{-}c_{1}^{2} ]   \\
m_{\nu_{\tau}} & = &  m_{0}[ 1 + \frac{1}{2}(\Delta_{+} + \Delta_{-}s_{1}^{2} ) + \frac{1}{2} 
(\Delta_{+} - \Delta_{-}s_{1}^{2})\sqrt{1 + \tan^{2}2\theta_{\nu} }\   ] \nonumber 
\end{eqnarray}
For the case C, the lepton mixing matrix and three neutrino masses are obtained as follows
\begin{equation}
C:\qquad U_{LEP} = U_{e}^{'\dagger}U_{e}^{\dagger}O_{\nu} = U_{e}^{'\dagger}\left( \begin{array}{ccc}
  c_{1} & -is_{1}c_{\nu} & -is_{1}s_{\nu}  \\
 -ic_{2}s_{1} & c_{1}c_{2}c_{\nu} + s_{2}s_{\nu} & c_{1}c_{2}s_{\nu}-s_{2}c_{\nu} \\
 -is_{1} s_{2} & c_{1}s_{2}c_{\nu}-c_{2}s_{\nu}  & c_{1}s_{2}s_{\nu}+c_{2}c_{\nu}  \\
\end{array} \right)
\end{equation}
and
\begin{eqnarray}
m_{\nu_{e}} & = &  m_{0}[ 1 + \Delta_{-}s_{1}^{2} ]  \nonumber \\
m_{\nu_{\mu}} & = &  m_{0}[ 1 + \frac{1}{2}(\Delta_{+} + \Delta_{-}c_{1}^{2} ) - 
\frac{1}{2} (\Delta_{+} - \Delta_{-}c_{1}^{2})\sqrt{1 + \tan^{2}2\theta_{\nu} } \  ]   \\
m_{\nu_{\tau}} & = &  m_{0}[ 1 + \frac{1}{2}(\Delta_{+} + \Delta_{-}c_{1}^{2} ) + \frac{1}{2} 
(\Delta_{+} - \Delta_{-}c_{1}^{2})\sqrt{1 + \tan^{2}2\theta_{\nu} }\   ] \nonumber 
\end{eqnarray}

   Before going to a detailed analysis, we would like to provide a brief summary of recent experiments. 
when the $\nu_{\mu}$ anomaly reported recently by Super-Kamiokande (SK) experiment\cite{SUPERK1} 
is interpreted as oscillation of $\nu_{\mu}\rightarrow \nu_{\tau}$ with nearly maximal mixing angle 
in a two flavor oscillation. The allowed range for mass-squared difference and 
mixing angle \cite{SUPERK1,BKS,FLMS} is  
\begin{equation}
\Delta m_{atm}^{2}=|\Delta m_{\mu\tau}^{2}| \simeq (0.5-6)\times 10^{-3} eV^{2}, 
\qquad \sin^{2}2\theta_{atm} > 0.82\  (90\%C.L.)
\end{equation}  
For three-flavor mixing case, the best fit to the sub-GeV, multi-GeV and upward-going muon data from SK  
is obtained at \cite{FLMS}
\begin{equation}
|\Delta m_{\mu\tau}^{2}| \simeq 2.5\times 10^{-3} eV^{2}, \qquad \sin^{2}\theta_{\mu\tau}=0.63, 
\qquad  \sin^{2}\theta_{e\tau}=0.14
\end{equation}  
The deficit in the solar neutrino experiments suggests the following best-fit solution from a global 
analysis \cite{BKS} 

 (1)  MSW\cite{MSW} small angle solution: $\Delta m_{sun}^{2} \simeq 5\times 10^{-6} eV^{2}, \qquad 
\sin^{2} 2\theta_{sun} \simeq 5.5\times 10^{-3}; $ \\
 (2) ``Just-so" vacuum\cite{JS} solution: $ \Delta m_{sun}^{2} 
\simeq 6.5\times 10^{-11} eV^{2}, \qquad \sin^{2} 2\theta_{sun} \simeq 0.75 $ 

although it allows three possible MSW solutions\footnote{It was shown in ref.\cite{BK} that 
if the low energy cross section for $He + p \rightarrow He + e^{+} + \nu_{e}$ reaction is 
about 20 times larger than the best (but uncertain) theoretical estimates, it allows 
three MSW solutions at $95\%$ C.L..} from rates only and a large range of mass-squared 
difference and mixing angle from spectrum only. For comparison, we list the allowed regions 
($99\%$ C.L.) from rates only (see Figs.(2) and (5) in ref. \cite{BKS}), 

(1) MSW small angle solution: $\Delta m_{sun}^{2} \simeq (3-10)\times 10^{-6} eV^{2},\ 
\sin^{2}2\theta_{sun} \simeq (1.5-10)\times 10^{-3}; $ \\
(2) MSW large angle solution: $\Delta m_{sun}^{2} \simeq (6-30)\times 10^{-5} eV^{2}, 
\qquad \sin^{2}2\theta_{sun} \simeq 0.4\sim 1; $ \\
(3) MSW Low solution: $\Delta m_{sun}^{2} \simeq (3.5-20)\times 10^{-8} eV^{2}, 
\qquad \sin^{2}2\theta_{sun} \simeq 0.8\sim 1; $ \\
(4) Vacuum Oscillation solution: $ \Delta m_{sun}^{2} 
\simeq (5-100)\times 10^{-11} eV^{2}, \qquad \sin^{2}2\theta_{sun} > 0.55 $ 

 We now turn to discuss possible interesting scenarios in the present model.

\section{Small-Large Mixing Scenario}

  It is easily seen that a small-large mixing scenario is most likely to occur 
in the case A. For simplicity, we may call this scenario as scenario (i). 
If the solar neutrino date is explained via $\nu_{e}\leftrightarrow \nu_{\mu}$ oscillation and 
the atmospheric neutrino data via $\nu_{\nu}\leftrightarrow \nu_{\tau}$ oscillation, 
the mass parameters are fixed to be
\begin{eqnarray}
& & \delta_{-} = \frac{\Delta m_{atm}^{2}}{4m_{0}^{2}(c_{2}^{2} -s_{2}^{2})} \nonumber \\
& & \delta_{+} = \frac{\Delta m_{atm}^{2}}{4m_{0}^{2}} \left(1 + 
\frac{\Delta m_{sun}^{2}}{2\Delta m_{atm}^{2}} \right) \nonumber 
\end{eqnarray} 
Nevertheless, within the present simple scheme the mixing elements $U_{ei}$ and $U_{ie}$ vanish. 
To obtain a realistic scheme, it is necessary to extend the present simple scheme and introduce new 
contributions or consider possible higher order corrections. 

    It is seen that as long as the mass scale 
$m_{0} >> \sqrt{\Delta m_{atm}^{2}}/2 \simeq (0.02\sim 0.04)$ eV, 
the three neutrino masses are almost degenerate. $m_{0}$ is mainly constrained from 
the neutrinoless double beta decay. In this scenario (i), it must satisfy\cite{DBD}  
\begin{equation}
m_{0} < 0.46 eV
\end{equation} 
The relation between the total neutrino mass $m(\nu)$ and the fraction $\Omega_{\nu}$ of 
critical density that neutrinos contribute is \cite{HDM}
\begin{equation}
\frac{\Omega_{\nu}}{\Omega_{m}} = 0.03 \frac{m(\nu)}{1eV} \left(\frac{0.6}{h}\right)^{2} 
\frac{1}{\Omega_{m}} \simeq 0.09 \frac{m_{0}}{1eV} \left(\frac{0.6}{h}\right)^{2} 
\frac{1}{\Omega_{m}}  
\end{equation}
with $h=0.5-0.8$ the expansion rate of the universe (Hubble constant $H_{0}$) in units of 100 km/s/Mpc. 
$\Omega_{m}$ is the fraction of critical density that matter contributes. In the second equility,
we have used the relation $m(\nu) \simeq 3 m_{0}$ for nearly degenerate neutrino mass.
Thus the upper bound implies that in the scenario (i) the fraction $\Omega_{\nu}/\Omega_{m} < 4\%$ 
for $\Omega_{m} = 1$ and $\Omega_{\nu}/\Omega_{m} < 10\%$ for $\Omega_{m} = 0.4$.

 \section{Nearly Bi-maximal Mixing Scenario} 

   The scenario (i) discussed in the previous section can be regarded as a two-flavor oscillation of 
a small-large mixing pattern since the mixing matrix element $(U_{LEP})_{13}$ is much smaller than 
other mixing matrix elements.  Another interesting two-flavor oscillation scenario 
occurs in a nearly bi-maximal mixing pattern. We will show that such a two-flavor oscillation 
is easily realized in the cases B and C  when $\tan^{2}2\theta_{\nu} << 1$. 
 
 By taking $\tan^{2}2\theta_{\nu} << 1$ and noting that $\sin^{2} 2\theta_{2} > 0.8 $ suggested 
from the recent atmospheric neutrino data\cite{SUPERK1}, we have, to a good approximation, 
the simple relations: $\Delta_{+} \simeq \Delta_{-} \simeq \delta_{+}$ and 
\begin{eqnarray}
& & B: \qquad \tan 2\theta_{\nu}\simeq 
2s_{1}\delta_{-}/\delta_{+}c_{1}^{2} << 1 \nonumber \\ 
& & C: \qquad \tan 2\theta_{\nu}\simeq 
2c_{1}\delta_{-}/\delta_{+}s_{1}^{2} << 1 
\end{eqnarray}
 Thus masses of the three neutrinos are simply given by 
\begin{eqnarray}
B:\qquad m_{\nu_{e}} & \simeq &  m_{0}[ 1 + \delta_{+}s_{1}^{2}  - 
\frac{\delta_{-}^{2}s_{1}^{2}}{\delta_{+}c_{1}^2} \  ] 
\nonumber \\
m_{\nu_{\mu}} & \simeq &  m_{0}[ 1 + \delta_{+}c_{1}^{2} ]  \nonumber  \\
m_{\nu_{\tau}} & \simeq &  m_{0}[ 1 + \delta_{+} + 
\frac{\delta_{-}^{2}s_{1}^{2}}{\delta_{+}c_{1}^2}  \   ] \nonumber \\
C:\qquad m_{\nu_{e}} & \simeq &  m_{0}[ 1 + \delta_{+}s_{1}^{2} ]  \\
 m_{\nu_{\mu}} & \simeq &  m_{0}[ 1 + \delta_{+}c_{1}^{2}  - 
\frac{\delta_{-}^{2}c_{1}^{2}}{\delta_{+}s_{1}^2} \  ]   \nonumber \\
m_{\nu_{\tau}} & \simeq &  m_{0}[ 1 + \delta_{+} + 
\frac{\delta_{-}^{2}c_{1}^{2}}{\delta_{+}s_{1}^2}  \   ] \nonumber 
\end{eqnarray}  
 From these masses,  one easily reads off the mass-squared differences 
\begin{eqnarray}
B: \qquad \Delta m_{\mu e}^{2} & = & m_{\nu_{\mu}}^{2} - m_{\nu_{e}}^{2} \simeq  
m_{0}^{2}\delta_{+}[c_{1}^{2}-s_{1}^{2} + \left(\frac{\delta_{-}s_{1}}{\delta_{+}c_{1}}\right)^{2} ] 
[2 + \delta_{+} ]
\nonumber \\
\Delta m_{\tau\mu}^{2} & = & m_{\nu_{\tau}}^{2} - m_{\nu_{\mu}}^{2} \simeq 
m_{0}^{2}\delta_{+}[s_{1}^{2} + \left(\frac{\delta_{-}s_{1}}{\delta_{+}c_{1}}\right)^{2} ] 
[2 + \delta_{+}(1 + c_{1}^{2}) ] \nonumber  \\
C:\qquad \Delta m_{\mu e}^{2} & = & m_{\nu_{\mu}}^{2} - m_{\nu_{e}}^{2} \simeq  
 m_{0}^{2}\delta_{+}[c_{1}^{2}-s_{1}^{2} -\left(\frac{\delta_{-}c_{1}}{\delta_{+}s_{1}}\right)^{2} ] 
[2 + \delta_{+} ] \\
\Delta m_{\tau\mu}^{2} & = & m_{\nu_{\tau}}^{2} - m_{\nu_{\mu}}^{2} \simeq 
m_{0}^{2}\delta_{+}[s_{1}^{2} + \left(\frac{\sqrt{2}\delta_{-}c_{1}}{\delta_{+}s_{1}}\right)^{2} ] 
[2 + \delta_{+}(1 + c_{1}^{2}) ] \nonumber 
\end{eqnarray} 

  To explain the atmospheric neutrino anomaly and the observed deficit of the solar neutrino 
fluxes in comparison with the solar neutrino fluxes computed from the solar 
standard model\cite{BP}, the required neutrino mass-squared 
difference $\Delta m_{\tau\mu}^{2}$ must satisfy\cite{SUPERK1,BKS,FLMS}
\begin{equation}
5\times 10^{-4} eV^{2} < |\Delta m_{\tau\mu}^{2}| < 6\times 10^{-3} eV^{2}
\end{equation}
and the mass-squared difference $\Delta m_{\mu e}^{2}$ shall fall into the range\cite{BKS}:
\begin{equation}
6\times 10^{-11} eV^{2} < \Delta m_{\mu e}^{2} < 2\times 10^{-5} eV^{2}
\end{equation} 
Here the larger and smaller values of $\Delta m_{\mu e}^{2}$ provide MSW\cite{MSW} 
and vacuum oscillation explanations for the solar neutrino puzzle respectively. It is seen that 
the ratio between the two mass-squared differences satisfy
$\Delta m_{\mu e}^{2}/\Delta m_{\tau\mu}^{2}< 0.04$. With this condition and $|\delta_{-}|<<|\delta_{+}|$, 
we then obtain from eq.(29) the following constraint on the mixing angle $\theta_{1}$ 
\begin{equation}
|c_{1}^{2}/s_{1}^{2} - 1| < 0.04
\end{equation}
Note that this constraint is independent of the mass scale $m_{0}$. 
With these constraints, we arrive at the following relations
\begin{equation} 
\frac{m''_{1}}{m'_{1}} \sim \sqrt{\frac{m_{e}}{m_{\mu}}} = 0.07, \qquad  
\frac{m''_{1}}{m_{1}} \sim \frac{\sqrt{m_{e}m_{\mu}}}{m_{\tau}} = 0.004
\end{equation}
Because of the smallness of the mixing angles in $U'_{e}$ and $\theta_{\nu}$, we may
conclude that the neutrino mixing between $\nu_{e}$ and $\nu_{\mu}$ is almost maximal
\begin{equation} 
\sin^{2}2\theta_{1} > 0.998
\end{equation} 
which may leave vacuum oscillation as the only viable explanation 
of the solar neutrino data as it can be seen from the analyses in \cite{MY}. 
This requires that $\sigma_{1} \simeq \sigma_{2}$ and
$m'_{0}\simeq m''_{0}$ which may need a fine-tuning if they are not ensured by symmetries.  

   With the above analyses, we may come to the conclusion that with two-flavor oscillation
and the hierarchical mass-squared differences $\Delta m_{\mu e}^{2}<<\Delta m_{\tau\mu}^{2}$,
both cases B and C lead to a nearly `bi-maximal' neutrino mixing pattern for 
the explanations of the solar and atmospheric neutrino flux anomalies. 

  The smallness of the mass-squared difference $\Delta m_{\mu e}^{2}$ implies that 
$\sin\theta_{\nu} < 0.001$ for $m_{0} \sim 2$ eV. To a good approximation, we may neglect the small 
mixing angle $\theta_{\nu}$ and the small mixing of order $m''_{1}/m_{1}$ in $U'_{e}$. 
Thus the CKM-type lepton mixing matrices for the cases B and C are simply given by 
\begin{eqnarray}
& & B:\qquad U_{LEP} \simeq
\left( \begin{array}{ccc}
  \frac{1}{\sqrt{2}}i & -\frac{1}{\sqrt{2}} & -i\sqrt{\frac{m_{e}}{m_{\mu}}}s_{2}  \\
  \frac{1}{\sqrt{2}}c_{2} & -\frac{1}{\sqrt{2}}c_{2}i & -s_{2} \\
 \frac{1}{\sqrt{2}}s_{2} & -\frac{1}{\sqrt{2}}s_{2}i & c_{2}  \\
\end{array} \right) \nonumber \\
& & C:\qquad  U_{LEP} \simeq
\left( \begin{array}{ccc}
  \frac{1}{\sqrt{2}} & -\frac{1}{\sqrt{2}}i & i\sqrt{\frac{m_{e}}{m_{\mu}}}s_{2}  \\
  -\frac{1}{\sqrt{2}}c_{2}i & \frac{1}{\sqrt{2}}c_{2} & -s_{2} \\
 -\frac{1}{\sqrt{2}}s_{2}i & \frac{1}{\sqrt{2}}s_{2} & c_{2}  \\
\end{array} \right) 
\end{eqnarray}
which arrive at the pattern suggested by Vissani\cite{FV} 
once neglecting the small mixing angle at the order of magnitude $\sqrt{m_{e}/m_{\mu}}s_{2}$.
When going back to the weak gauge and charged-lepton mass basis, 
we find that the neutrino mass matrices for the cases B and C have the following simple forms   
\begin{eqnarray}
& & B:\qquad M_{\nu} \simeq m_{0}\left( \begin{array}{ccc}
 -\frac{m_{e}}{m_{\mu}}s_{2}^{2} & ic_{2} & is_{2}  \\
   ic_{2} & s_{2}^{2} &  - c_{2}s_{2}  \\
  is_{2} &  - c_{2}s_{2} &  c_{2}^{2}  \\ 
\end{array} \right) \nonumber \\
& & C:\qquad M_{\nu} \simeq m_{0}\left( \begin{array}{ccc}
 -\frac{m_{e}}{m_{\mu}}s_{2}^{2} & -ic_{2} & -is_{2}  \\
   -ic_{2} & s_{2}^{2} &  - c_{2}s_{2}  \\
  -is_{2} &  - c_{2}s_{2} &  c_{2}^{2}  \\ 
\end{array} \right)
\end{eqnarray}

 Suggested by the recent atmospheric neutrino data,
we are motivated to consider two particular interesting cases: Firstly, setting the 
vacuum expectation values to be $\sigma_{3}^{2}
= \sigma_{1}^{2} + \sigma_{2}^{2}$ and $\sigma_{1} = \sigma_{2}$, 
namely, $s_{1}=s_{2}=1/\sqrt{2}$ ($\sin^{2}2\theta_{1} =\sin^{2}2\theta_{2}=1$), 
we then obtain a realistic bi-maximal mixing pattern with 
a maximal CP-violating phase. Explicitly, the neutrino mass and mixing 
matrices read 
\begin{eqnarray}
& & B:\qquad M_{\nu} \simeq m_{0}\left( \begin{array}{ccc}
  -0.002 & \frac{1}{\sqrt{2}}i & \frac{1}{\sqrt{2}}i  \\
   \frac{1}{\sqrt{2}}i & \frac{1}{2} &  - \frac{1}{2}  \\
  \frac{1}{\sqrt{2}}i &  - \frac{1}{2} &  \frac{1}{2}  \\ 
\end{array} \right) \nonumber \\
& & C:\qquad M_{\nu} \simeq m_{0}\left( \begin{array}{ccc}
  -0.002 & -\frac{1}{\sqrt{2}}i & -\frac{1}{\sqrt{2}}i \\
   -\frac{1}{\sqrt{2}}i & \frac{1}{2} &  - \frac{1}{2}  \\
  -\frac{1}{\sqrt{2}}i &  - \frac{1}{2} &  \frac{1}{2}  \\ 
\end{array} \right)
\end{eqnarray}
and 
\begin{eqnarray}
& & B:\qquad U_{LEP} \simeq
\left( \begin{array}{ccc}
  \frac{1}{\sqrt{2}}i & -\frac{1}{\sqrt{2}} & -0.05i  \\
  \frac{1}{2} & -\frac{1}{2}i & -\frac{1}{\sqrt{2}} \\
 \frac{1}{2} & -\frac{1}{2}i & \frac{1}{\sqrt{2}}  \\
\end{array} \right)  \nonumber \\
& & C:\qquad U_{LEP} \simeq
\left( \begin{array}{ccc}
  \frac{1}{\sqrt{2}} & -\frac{1}{\sqrt{2}}i & 0.05i  \\
  -\frac{1}{2}i & \frac{1}{2} & -\frac{1}{\sqrt{2}} \\
 -\frac{1}{2}i & \frac{1}{2} & \frac{1}{\sqrt{2}}  \\
\end{array} \right) 
\end{eqnarray}
when neglecting the small mixing angle at the order of magnitude $\sqrt{m_{e}/m_{\mu}}$, 
we then yield the pattern suggested by  Georgi and Glashow\cite{GG}. 

Secondly, setting the three vacuum expectation values $\sigma_{i}$ (i=1,2,3) 
to be democratic, i.e., $\sigma_{1} = \sigma_{2}=\sigma_{3}$,
hence $s_{1}=1/\sqrt{2}$ and $s_{2}=\sqrt{2/3}$ 
($\sin^{2}2\theta_{1}=1$ and $\sin^{2}2\theta_{2}= 8/9$), we 
then arrive at a realistic democratic mixing pattern with a maximal
CP-violating phase. The explicit neutrino mass and mixing matrices 
are given by 
\begin{eqnarray}
& & B:\qquad M_{\nu} \simeq m_{0}\left( \begin{array}{ccc}
  -0.003 & \frac{1}{\sqrt{3}}i & \frac{2}{\sqrt{6}}i  \\
   \frac{1}{\sqrt{3}}i & \frac{2}{3} &  - \frac{\sqrt{2}}{3}  \\
  \frac{2}{\sqrt{6}}i  &  - \frac{\sqrt{2}}{3} &  \frac{1}{3}  \\ 
\end{array} \right)  \nonumber  \\
& & C:\qquad M_{\nu} \simeq m_{0}\left( \begin{array}{ccc}
  -0.003 & -\frac{1}{\sqrt{3}}i & -\frac{2}{\sqrt{6}}i  \\
   -\frac{1}{\sqrt{3}}i & \frac{2}{3} &  - \frac{\sqrt{2}}{3}  \\
  -\frac{2}{\sqrt{6}}i  &  - \frac{\sqrt{2}}{3} &  \frac{1}{3}  \\ 
\end{array} \right)  
\end{eqnarray}
and 
\begin{eqnarray}
& & B: \qquad U_{LEP} \simeq 
\left( \begin{array}{ccc}
  \frac{1}{\sqrt{2}}i & -\frac{1}{\sqrt{2}} & -0.057i  \\
  \frac{1}{\sqrt{6}} & -\frac{1}{\sqrt{6}}i & - \frac{2}{\sqrt{6}} \\
 \frac{1}{\sqrt{3}} & -\frac{1}{\sqrt{3}}i & \frac{1}{\sqrt{3}}  \\
\end{array} \right)  \nonumber \\
& & C: \qquad U_{LEP} \simeq 
\left( \begin{array}{ccc}
  \frac{1}{\sqrt{2}} & -\frac{1}{\sqrt{2}}i & 0.057i  \\
  -\frac{1}{\sqrt{6}}i & \frac{1}{\sqrt{6}} & - \frac{2}{\sqrt{6}} \\
 -\frac{1}{\sqrt{3}}i & \frac{1}{\sqrt{3}} & \frac{1}{\sqrt{3}}  \\
\end{array} \right) 
\end{eqnarray}
when further neglecting the small mixing angle at the order of magnitude 
$\sqrt{m_{e}/m_{\mu}}$, we obtain a similar form provided by Mohapatra\cite{RABI}.   

    We may call the above scenario as `scenario (ii)'. 
From the hierarchical feature in $\Delta m^{2}$, i.e., $\Delta m_{\mu e}^{2}<<
\Delta m_{\tau\mu}^{2}\simeq \Delta m_{\tau e}^{2}$, 
formulae for the oscillation probabilities in the scenario (ii) can be greatly simplified to be
\begin{eqnarray}
& & P_{\nu_{e}\rightarrow \nu_{e}}|_{solar} \simeq  1 - 
4|U_{e1}|^{2}|U_{e2}|^{2} \sin^{2}(\frac{\Delta m_{\mu e}^{2}L}{4E}) \nonumber \\
& & P_{\nu_{\mu}\rightarrow \nu_{\mu}}|_{atmospheric} \simeq 1 -4(1-|U_{\mu 3}|^{2})|U_{\mu 3}|^{2}
\sin^{2}(\frac{\Delta m_{\tau\mu}^{2}L}{4E}) \\
& & P_{\nu_{\beta}\rightarrow \nu_{\alpha}} \simeq 4|U_{\beta 3}|^{2}|U_{\alpha 3}|^{2}
\sin^{2}(\frac{\Delta m_{\tau\mu}^{2}L}{4E}) \nonumber 
\end{eqnarray}
and 
\begin{equation}
\frac{P_{\nu_{\mu}\rightarrow \nu_{e}}}{ P_{\nu_{\mu}\rightarrow \nu_{\tau}}}|_{atmospheric}
\simeq \frac{|U_{e 3}|^{2}}{|U_{\tau 3}|^{2}} << 1
\end{equation}
This may present the simplest scheme for reconciling both 
solar and atmospheric neutrino fluxes via oscillations of three neutrinos. 

 Furthermore, the resulting `bi-maximal' neutrino mixing pattern 
allows the three neutrino masses to be nearly degenerate and large enough for hot dark matter
without conflict with the current data on neutrinoless double beta decay. 
This is because the failure of detecting neutrinoless double beta decay provide 
bounds on an effective electron neutrino mass $<m_{\nu_{e}}> = \sum_{i} 
m_{\nu_{i}}(U_{LEP})_{ei}^{2}  < 0.46$ eV\cite{DBD}. In the present scenario, we have 
$|(M_{\nu})_{ee}|= <m_{\nu_{e}}> \simeq (0.002-0.03)m_{0} $. It shows that the mass scale 
can be as large as $m_{0} \sim 15$ eV. From the relation between the total neutrino 
mass $m(\nu)$ and the fraction $\Omega_{\nu}$ of 
critical density that neutrinos contribute, we find that for $m_{0} \sim 2$ eV and $h=0.6$
the fraction $\Omega_{\nu} \simeq 18 \%$ for $\Omega_{m} = 1$.

\section{Three-Flavor Oscillation Scenario}

  As a general case, it is interesting to consider three-flavor oscillation which may occur 
in the cases B and C when the mixing angle $\theta_{\nu}$ is not small. As shown in ref.\cite{FLMS}, 
nonzero values of $U_{e3}^{2}$ may explain part of the electron excess in the Sub-GeV and Multi-GeV samples
and can also contribute to distort the zenith distributions.  Direct constraints on $U_{e3}^{2}$ arise from 
the CHOOZ experiment. The CHOOZ data can provide a strong constraint on the higher range of 
$\Delta m_{atm}^{2}$, but its role decreases rapidly as $\Delta m_{atm}^{2}$ goes down to $10^{-3}$ eV$^{2}$. 
It was pointed out in ref.\cite{FLMS} that within the framework of three-flavor oscillation, 
for $\Delta m_{atm}^{2}$ close to $10^{-3}$ (or slightly below ) $10^{-3}$ eV$^{2}$, 
the allowed region at $99\%$ C.L. by the SK and CHOOZ as well as solar neutrino data could be
\begin{equation}
U_{e2}^{2} = \frac{8}{25}, \qquad U_{e3}^{2} = \frac{1}{5}, \qquad U_{\mu3}^{2} = \frac{4}{5}
\end{equation} 
If using this set of mixing as input, we find that the three mixing angles $\theta_{1}$, 
$\theta_{\nu}$ and $\theta_{2}$ for the cases B and C in the present model are fixed to be
\begin{eqnarray}
& & B:\qquad  \tan^{2}\theta_{1} = \frac{8}{17}, \qquad \tan 2\theta_{\nu} = -\frac{4\sqrt{15}}{7}, 
\qquad \tan 2\theta_{2} = \frac{13}{2\sqrt{30}}  \nonumber \\
& & C:\qquad  \tan^{2}\theta_{1} = \frac{13}{12}, \qquad \tan 2\theta_{\nu} = -\frac{4\sqrt{10}}{3}, 
\qquad \tan 2\theta_{2} = \frac{91}{2\sqrt{5070}} 
\end{eqnarray} 
Thus the corresponding mixing matrices for the cases B and C are 
\begin{eqnarray}
& & B: \qquad U_{LEP} \simeq 
U_{e}^{'\dagger}\left( \begin{array}{ccc}
  -0.693i & -0.566 & 0.447i  \\
  -0.202 & -0.748 i & 0.632 \\
 -0.692 & -0.348 i & -0.632  \\
\end{array} \right)  \nonumber \\
& & C: \qquad U_{LEP} \simeq 
U_{e}^{'\dagger}\left( \begin{array}{ccc}
  0.693  & 0.566i & -0.447i  \\
  -0.692 i & -0.348 & 0.632 \\
  -0.202 i & -0.748 & -0.632 \\
\end{array} \right) 
\end{eqnarray}
where one can neglect the small mixing in $U_{e}^{'\dagger}$ since the largest mixing in 
$U_{e}^{'\dagger}$ is $s_{e}\simeq m''_{1}\sin 2\theta_{1}c_{2}/m_{\mu} \simeq 0.049$ 
as $m''_{1} \simeq 5.56 $eV for the case C and  $s_{e} \simeq -0.011$ as $m''_{1} \simeq -1.36$ eV 
for the case B. This mixing pattern may be simply called as scenario (iii). 
 
It is interesting to note that once the three angles are given, 
the ratio of the two mass-squared differnces 
$\Delta m_{atm}^{2}$ and $\Delta m_{sun}^{2}$ is completely determined in the present model to be
\begin{equation}
B:\qquad \frac{|\Delta m_{sun}^{2}|}{|\Delta m_{atm}^{2}|} \simeq 0.036; \qquad C:\qquad 
\frac{|\Delta m_{sun}^{2}|}{|\Delta m_{atm}^{2}|} \simeq 0.09
\end{equation}
For $\Delta m_{atm}^{2} \simeq 1.0\times 10^{-3}$ eV$^{2}$, the resulting mass-squared difference 
$\Delta m_{sun}^{2}\sim 0.9\times 10^{-4}$ eV$^{2}$ for the case C and 
is found to be consistent with the experimental data for the allowed region at $99\%$ C.L., 
while the numerical result $\Delta m_{sun}^{2}\sim 0.36\times 10^{-4}$ eV$^{2}$ for the case B 
is quite near the low bound allowed from the solar neutrino data\cite{FLM} at $99\%$ C.L..

As $c_{1}^{2}-s_{1}^{2} = 1/12$ for the case C, the degenerate neutrino mass can be large enough 
to play a significant cosmological role. For the case B, the mass scale $m_{0}$ is bounded to 
be $m_{0} < 0.46 eV/(c_{1}^{2}-s_{1}^{2})=1.28$ eV from the neutrinoless double beta decay. 
Thus the fraction $\Omega_{\nu}/\Omega_{m}$ is bounded to be $\Omega_{\nu}/\Omega_{m} < (12-29)\%$ 
for $\Omega_{m} = (1.0-0.4)$ and $h=0.6$. 
 
\section{Conclusions and Remarks}

  Starting from the effective lagrangian in eq.(1) with SO(3)$_{F}\times$ SU(2)$_{L}\times$ U(1)$_{Y}$ 
symmetry, we have investigated three possible interesting scenarios: small-large mixing scenario 
(the scenario (i)), nearly bi-maximal mixing scenario (the scenario (ii)) and three-flavor oscillation 
scenario (the scenario (iii)) to reconcile the solar and atmospheric neutrino anomalies 
when LSND results\cite{LSND} are not considered. This is because including the LSND results, 
it likely needs to introduce a sterile neutrino\cite{STERILE}. 
The three scenarios are found in connection with the three possible assignments (i.e., cases A, B and C ) 
of the maximal CP-violating phase to one of three real amplitude fields of the SO(3) triplet 
scalar after spontaneous symmetry breaking of the SO(3) gauge symmetry. The scenario (i) may only 
arise from the case A, but within the present simple model, one cannot yet obtain the realistic small 
mixing angle needed for MSW solution of solar neutrino. In contrast, it is of interest to see that both 
cases B and C can easily lead to the scenario (ii). The scenario (iii) 
may also be derived from the cases B and C to accomodate the solar and atmospheric 
neutrino data for the allowed region at $99\%$ C.L., while the parameter space for the case B is 
limited to be quite small even if for the allowed region at $99\%$ C.L. and this scenario seems 
rather unlikely in the case B. We therefore conclude that within the present simple model the scenario 
(ii) should be more reliable to be constructed to reconcile both solar and atmospheric neutrino data.
It would be interesting to investigate the vacuum structures of spontaneous SO(3) symmetry breaking 
to understand the real case. For that, it may be useful to work in 
a supersymmetric SO(3)$_{F}\times$ SU(2)$_{L}\times$ U(1)$_{Y}$ gauge theory and regard 
the complex triplet scalar fields as superfields\cite{BHKR}.

 {\bf Ackowledgments}: The author would like to thank J. Bahcall for bring his attention on the 
paper\cite{BK} and E. Lisi and S. Pakvasa for kind remarks. He would also like to thank 
C.S. Lam for fruitful communication. This work was supported in part by 
the NSF of China under the grant No. 19625514.



\begin{thebibliography}{99}
\bibitem{SUPERK1} Super-Kamiokande Collaboration, Y.Fukuda {\it et al.}, Phys. Rev. Lett. 
{\bf 81}, 1562 (1998).
\bibitem{SUPERK2} Super-Kamiokande Collaboration, Y.Fukuda {\it et al.}, Phys. Rev. Lett. 
{\bf 81}, 1158 (1998).
\bibitem{SOLAR} The history and progresses on solar neutrinos may been found 
from the J.N. Bahcall internet page, at www.sns.ias.edu/ \~  \  jnb..
\bibitem{BP} J.N. Bahcall and M.H. Pinsonneault, Rev. Mod. Phys. {\bf 67}, 781 (1995) (`BP95'); 
 J.N. Bahcall, S. Basu and M.H. Pinsonneault, Phys. Lett. {\bf B433}, 1 (1998) (`BP98').
\bibitem{YLWU} Y.L. Wu, hep-ph/9810491, Phys. Rev. D, 1999 (in press); hep-ph/9901245,
Euro. Phys. J. C, 1999 (in press).
\bibitem{BMAX}  V. Barger, S. Pakvasa, T. Weiler and K. Whisnant, 
Phys. Lett. {\bf B437}, 107 (1998); \\ 
Y. Nomura and T. Yanagida, hep-ph/9807325; \\
R.N. Mohapatra and S. Nussinov, hep-ph/9809415; \\
M.Fukugita, M. Tanimoto and T. Yanagida, hep-ph/9809554; \\
C. Jarlskog, M. Matsuda, S. Skadhauge and M. Tanimoto, hep-ph/9812282.
\bibitem{DEM}  H. Fritzsch and Z.Z. Xing, Phys. Lett. {\bf B 372} 265 (1996); \\  
 M.Fukugita, M. Tanimoto and T. Yanagida, Phys. Rev. {\bf D57}, 4429 (1998); \\
 Y. Koide, Phys. Rev. {\bf D 39} 1391 (1989); \\ 
R.N. Mohapatra and S. Nussinov, hep-ph/9808301; \\
M. Tanimoto, hep-ph/9807283.
Y. Koide, Phys. Rev. {\bf D 39} 1391 (1989).
\bibitem{CM} C. Carone and M. Sher, Phys. Lett. {\bf B420}, 83 (1998).
\bibitem{MA} E. Ma, hep-ph/9812344.
\bibitem{CW} C. Wetterich, hep-ph/9812426.
\bibitem{BHKR} R. Barbieri, L.J. Hall, G.L. Kane and G.G. Ross, hep-ph/9901228.
\bibitem{BKS} J.N. Bahcall, P. Krastev and A.Yu. Smirnov, Phys. Rev. {\bf D58}, 096016 (1998).
\bibitem{FLMS} G.L. Fogli, E. Lisi, A. Marrone and G. Scioscia, BARI-TH/309-98, 
hep-ph/9808205.
\bibitem{MSW} L. Wolfenstein, Phys. Rev. {\bf D17} 2369 (1978); \\
S.P. Mikheyev and A. Yu. Smirnov, Yad.Fiz. {\bf 42} (1985) 1441 
[Sov. J. Nucl. Phys. {42} (1985) 913].
\bibitem{JS} V. Barger, K. Whisnant and R.J.N Phillips, Phys. Rev. {\bf D24} 538 (1981); \\
S.L. Glashow and L.M. Krauss, Phys. Lett. {\bf B190} (199 (1987).
\bibitem{BK} J.N. Bahcall and P.I. Krastev, Phys. Lett. {\bf B436}, 243 (1998).
\bibitem{DBD} Heideberg-Moscow collaboration, M. G\"{u}nther {\it et al.}, 
Phys. Rev. {D55}, 54 (1997);  L. Baudis {\it et al.}, Phys. Lett. {\bf B407} 219 (1997); 
Klapdor-Kleingrothaus, hep-ex/9802007.
\bibitem{HDM}J.R. Primack and M. A. K. Gross, astro-ph/9810204, to be published in the proceedings of
    the Xth Rencontres de Blois, "The Birth of Galaxies", 28 June - 4 July 1998; 
 J.R. Primack, J. Holtzman, A. Klypin and D.O. Caldwell, 
Phys. Rev. Lett. {\bf 74} 2160 (1995).
\bibitem{MY} H. Minakata and O. Yasuda, Nucl. Phys. {\bf B523}, 597 (1998); 
ibid., Phys. Rev. {\bf D56}, 1692 (1997).
\bibitem{FV} F. Vissani, hep-ph/9708483. 
\bibitem{GG}H. Georgi and S.L. Glashow, HUTP-98/A060, hep-ph/9808293.
\bibitem{RABI} R.N. Mohapatra, hep-ph/9808284.
\bibitem{FLM} G.L. Fogli, E. Lisi, A. Marrone, Phys. Rev. {\bf D54}, 2048 (1996).
\bibitem{LSND} LSND collaboration, C. Athanassopoulos et al., Phys. Rev. Lett. 
{\bf 75}, 365 (1995); ibid. {\bf 77}, 3082 (1996). 
\bibitem{STERILE} D.O. Caldwell and R.N. Mohapatra, Phys. Rev. 
{\bf D48} 3259 (1993); \\ J.T. Peltoniemi and J.W.F. Valle, 
Nucl. Phys. {B406} 409 (1993); \\
 R. Foot and R.R. Volkas, Phys. Rev. 
{\bf D52} 6595 (1995);  \\ K.C. Chou and Y.L. Wu,
 Nucl. Phys. B (Proc. Suppl.) {\bf 52A} (1997) 159-163, hep-ph/9610300;  
 hep-ph/9708201; \\  
 Zhou Guang-Zhao (K.C. Chou) and Wu Yue-Liang (Y.L. WU) 
 Science in China, 39A, 65 (1996), hep-ph/9508402; \\ 
S.M. Bilenky, C. Giunti and W. Grimus, Eur. Phys. J. 
 {\bf C1} 247 (1998); \\  V. Barger, K. Whisnant and T.J. Weiler, 
 Phys. Lett. {\bf B427}, 97 (1998); \\ S.C. Gibbons, 
  R.N. Mohapatra, S. Nandi and A. Raychaudhuri, Phys. Lett. {\bf B430}, 296 (1998), 
hep-ph/9803299; \\ V. Barger, S. Pakvasa, T.J. Weiler and K. Whisnant, 
Phys. Rev. {\bf D58}, 093016 (1998); \\
A. S. Joshipura and A. Yu. Smirnov, Phys. Lett. {\bf B439}, 103 (1998); \\
L. Hall and N. Weiner, hep-ph/9811299.
\end{thebibliography}
\end{document}